\documentclass[aps,reprint,amsmath,amssymb,aps,prx,superscriptaddress]{revtex4-2}

\usepackage[english]{babel} 
\usepackage{graphicx}
\usepackage{dcolumn}
\usepackage{bm}
\usepackage[dvipsnames]{xcolor}

\usepackage{mathtools} 
\usepackage{graphicx}
\usepackage{epstopdf}
\usepackage{amsmath}
\usepackage{dcolumn}
\usepackage{bm}
\usepackage{bbold}
\usepackage{amssymb}
\usepackage{mathrsfs}

\usepackage{amsfonts}
\usepackage{braket}
\usepackage{tabularx}
\usepackage{setspace}
\usepackage[caption=false]{subfig}

\usepackage{float}
\usepackage{array}
\usepackage[dvipsnames]{xcolor}
\newcolumntype{L}[1]{>{\raggedright\let\newline\\\arraybackslash\hspace{0pt}}m{#1}}
\newcolumntype{C}[1]{>{\centering\let\newline\\\arraybackslash\hspace{0pt}}m{#1}}
\newcolumntype{R}[1]{>{\raggedleft\let\newline\\\arraybackslash\hspace{0pt}}m{#1}}

\usepackage[makeroom]{cancel}

\usepackage{wrapfig}

\usepackage{hyperref}
\hypersetup{
    colorlinks=true,
    citecolor=black,
    linkcolor=black,
    filecolor=black,      
    urlcolor=black,
    pdftitle={Overleaf Example},
    pdfpagemode=FullScreen,
}

\begin{document}

\title[]{Electromagnetic Helicity in Twisted Cavity Resonators}	

\author{E. C. I. Paterson} 
\affiliation{Quantum Technologies and Dark Matter Labs, Department of Physics,  University of Western Australia, 35 Stirling Hwy, 6009 Crawley, Western Australia.}

\author{J. Bourhill}
\affiliation{Quantum Technologies and Dark Matter Labs, Department of Physics,  University of Western Australia, 35 Stirling Hwy, 6009 Crawley, Western Australia.}

\author{M. E. Tobar}
\affiliation{Quantum Technologies and Dark Matter Labs,  Department of Physics,  University of Western Australia, 35 Stirling Hwy, 6009 Crawley, Western Australia.}

\author{M. Goryachev}
\affiliation{Quantum Technologies and Dark Matter Labs,  Department of Physics,  University of Western Australia, 35 Stirling Hwy, 6009 Crawley, Western Australia.}

\date{\today}
	
\begin{abstract}
\noindent 
Through left- or right-handed twisting, we investigate the impact of mirror-asymmetry (chirality) of 
the conducting boundary conditions of an equilaterial triangular cross-section electromagnetic resonator.
We observe the generation of eigenmodes with non-zero electromagnetic helicity as a result of the coupling of near degenerate TE$_{11(p+1)}$ and TM$_{11p}$ 
modes. This can be interpreted as an emergence of magneto-electric coupling, which in turn produces a measurable shift in resonant mode frequency as a function of twist angle.
We show that this coupling mechanism is equivalent to introducing a non-zero chirality material parameter $\kappa_\text{eff}$ or axion field $\theta_{\text{eff}}$ to the medium. 
Our findings demonstrate the potential for real-time, macroscopic 
manipulation of electromagnetic helicity. 

\end{abstract}

\maketitle

\noindent Chiral (handed) electromagnetic radiation is distinguished by its left- or right-handed polarization, which arises from the intrinsic angular momentum of the electromagnetic field. The unique light-matter interactions exhibited by chiral radiation at the quantum level have garnered significant interest across various fields, including material 
science~\cite{Zhaofeng_Chiral_Metamaterials}, nanophotonics~\cite{Tang_Optical_Chirality,Zhao_Enhanced_Chiroptical,huang_spin-preserving_2020}, spectroscopy~\cite{Zhao_Enhanced_Chiroptical}, and quantum information processing~\cite{jonathan_Quantum,You_Quantum}. 
These interactions offer numerous applications, from probing molecular chirality, with applications for detecting biomarkers related to neurodegenerative diseases ~\cite{Wang2009ChiralMS}, to advancing optical communication technologies~\cite{Jia_Probing_Molecular_Chirality, MacKenzie_Chiral_Twists} and sensitivity to dark matter candidates \cite{Bourhill_twisted_anyon_cavity_2023}. 

Electromagnetic helicity, $\mathscr{H}$, is derived by performing a projection measurement of a complex electromagnetic state vector{'}s spin onto its linear momentum \cite{Alpeggiani_Electromagnetic_2018,PhysRevLett.113.033601,Martinez-Romeu:24,Bliokh_2013}, in fitting with the first-principles definition of the term {``}helicity{''}. The sign of this projection value will be opposite for different handed helicities and the expectation value of this operator will effectively measure the local, time averaged helicity density. It has been shown that the result of this projection is intrinsically linked to the dual transformation of the electromagnetic state, effectively rotating the electric and magnetic properties of the medium. This results in the effective mixing of electric and magnetic fields and so electromagnetic helicity will necessarily produce an electro-magnetic coupling. Most chiroptical phenomena, which produces electromagnetic helicity, occur as surface states, at optical frequencies, or due to complex meta-structures~\cite{Liu:2014ur,Khanikaev:2013vy,Goryachev2016,Bliokh_Topological_2019, huang_spin-preserving_2020,single_handedness_chiral_2022,chiral-mirrors-2015,chiral-light-twisted-fabry-2024,helicity-preserving-optical-2020,toward-molecular-chiral-polaritons-Baranov,effective-chiral-response-2024}. 

\begin{figure}[t!]
    \centering
    \includegraphics[width=0.95\columnwidth]{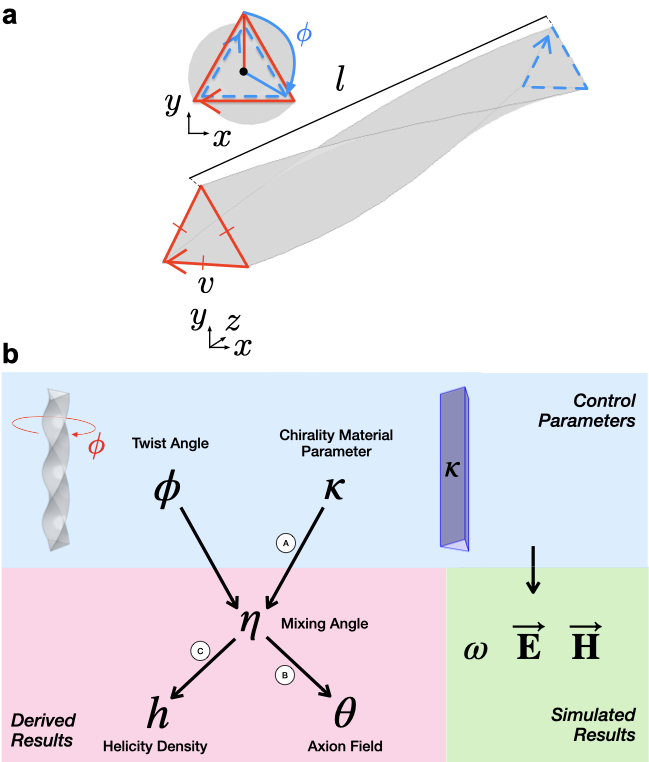}
    \caption{(a) The geometry of the twisted triangular cavity resonator.
    (b) A high-level diagram illustrating the mechanism for generating helical modes in a twisted triangular cavity resonator. By adjusting one of the two control 
    parameters; either twist angle, $\phi$, or the chirality parameter, $\kappa$, across the cavity volume in an FEM simulation, 
    a magneto-electric coupling is introduced within the cavity. This coupling results in a mixing angle $\eta$ between the electric field, $\vec{\mathbf{E}}$, and the magnetic field, 
    $\vec{\mathbf{H}}$, of the resonant electromagnetic radiation, leading to the generation of chiral electromagnetic modes with non-zero helicity density, $h$, and axion field, 
    $\theta$, both of which can be derived from the FEM simulated results. The circled labels A, B and C indicate the relationships between parameters, given by equations~\eqref{eq:kappa_eta}, \eqref{eq:mixing_angle} and \eqref{eq:eta_h} respectively.}
    \label{fig:EH_Relationship}
\end{figure}

We show that electromagnetic cavity resonators with twisted electrically conducting boundary conditions generate helical modes in vacuo without the need for engineered materials. We investigate a cavity with an equilateral triangular cross-section and electrically conducting boundary conditions at the end faces. The mirror symmetry of this resonator can be broken by introducing an arbitrary twist angle $\phi$ perpendicular to the central axis of the cavity, resulting in the chiral geometry shown in Fig.~\ref{fig:EH_Relationship}(a). A right-handed twist corresponds to $\phi > 0$, while a left-handed twist corresponds to $\phi < 0$. The breaking of spatial symmetry generates a magneto-electric coupling that mixes the near degenerate transverse electric (TE) and transverse magnetic (TM) modes, generating eigenmodes with non-zero $\mathscr{H}$.

Chirality can also be defined for materials, where the chirality of a material affects how chiral electromagnetic radiation interacts with it and is measured by the chirality material parameter $\kappa$. Material chirality is related to observable effects like optical rotation and circular dichroism~\cite{Multunas_Circular_Dichroism_Crystals}. We demonstrate the equivalency of twisting the resonator to introducing some effective, uniform, non-zero $\kappa$ over the cavity volume, which we refer to as $\kappa_\text{eff}$. We demonstrate the generation of a $\kappa_\text{eff}$ in free space that is two orders of magnitude larger than the bi-isotropoic effect exhibited by the material with the strongest known magneto-electric effect - Cr$_2$O$_3$~\cite{Hehl_Relativistic_2008}. 

Equivalent to $\kappa_\text{eff}$, the magneto-electric coupling can be thought to introduce an effective non-zero axion field $\theta_{\text{eff}}$, which is a pseudoscalar field hypothesised to solve the strong charge-parity (CP) problem~\cite{PQ1977,PhysRevD.16.1791,Weinberg1978,Wilczek1978} in quantum chromodynamics (QCD)~\cite{Peccei2006}. The strong CP-problem involves a CP-violating term theoretically allowed in the strong interaction not being observed in experiments. Introduced by the Peccei-Quinn mechanism, the axion field dynamically cancels out this CP-violating term, resolving the discrepancy.When quantized, the axion field gives rise to a particle called the axion, which is a strong candidate for dark matter. The generation of an axion field provides a method for a highly sensitive search for dark matter~\cite{Bourhill_twisted_anyon_cavity_2023}.

We demonstrate that regardless of the origin: twisting, chiral media, or axion field, the resulting mode mixing leads to a measurable shift in resonant frequency $\Delta\omega/\omega$ as a function of twist angle, $\phi$. Further, we demonstrate the inter-relationship between $\Delta\omega/\omega$, $\mathscr{H}$, the all-volume effective chirality parameter $\kappa_\text{eff}$ and the effective axion field $\theta_{\text{eff}}$ of the cavity mode. These relationships are summarised in Fig.~\ref{fig:EH_Relationship}(b). 

Ultimately, the improved understanding of these systems paves the way for optimal design and modelling of useful devices with the potential to generate a unique state of helical electromagnetic resonance within a macroscopic free-space volume. As has been hinted above, such devices may find use as molecular or solid-state chirality sensors, molecular sorters/filters, and sensors for fundamental physics such as dark-matter or gravity waves \cite{sym14102165}.

\noindent Total resonant helicity $\mathscr{H}_i$ for a given mode, $i$, can be formally defined for a monochromatic electromagnetic field in a resonant mode as the integral of the helicity density, $h_i$ 
over the cavity volume, $V$, expressed as: 
\begin{equation}
    h_i(\vec{r})=2\text{Im}\left[\vec{\mathbf{e}}_i(\vec{r})\cdot\vec{\mathbf{h}}_i^*(\vec{r})\right]=\frac{2\text{Im}\left[\vec{\mathbf{E}}_i(\vec{r})\cdot{\vec{\mathbf{H}}_i^*(\vec{r})}\right]}{V\mathcal{E}\mathcal{H}}
\label{eq:local_helicity}
\end{equation}
where $\mathcal{E}$ and $\mathcal{H}$ are real constants, $\vec{\mathbf{E}}_i(\vec{r})=\mathcal{E} \vec{\mathbf{e}}_i(\vec{r})$ and $\vec{\mathbf{H}}_i(\vec{r}) =\mathcal{H}\vec{\mathbf{h}}_i(\vec{r})$ are the electric and magnetic vector fields of the mode, respectively, and $\vec{\mathbf{e}}_i(\vec{r})$ and $\vec{\mathbf{h}}_i(\vec{r})$ are the normalised position dependent eigenvectors such that $\frac{1}{V}\int \vec{\mathbf{e}}_i(\vec{r})^*\cdot\vec{\mathbf{e}}_i(\vec{r})dV=\frac{1}{V}\int\vec{\mathbf{h}}_i(\vec{r})^*\cdot\vec{\mathbf{h}}_i(\vec{r})dV=1.$  Thus, the total mode helicity may be written as:
\begin{equation}
    \mathscr{H}_i=\int h_i dV. \label{eq:helicity}
\end{equation}
While $\mathscr{H}_i$ offers a quantitative measure of the global chirality of the electromagnetic radiation in the resonant mode, $h_i$ provides insight into the chirality of the radiation at specific points within the cavity. It is easy to show that $\mathcal{E}=\sqrt{\frac{1}{V}\int |\vec{\mathbf{E}}_i(\vec{r})|^2 dV}$ and $\mathcal{H}=\sqrt{\frac{1}{V}\int |\vec{\mathbf{H}}_i(\vec{r})|^2 dV}$, making (\ref{eq:local_helicity}) and (\ref{eq:helicity}) consistent with other definitions of helicity \cite{Bourhill_twisted_anyon_cavity_2023}.

\begin{figure}[b]
    \includegraphics[width=1.0\columnwidth]{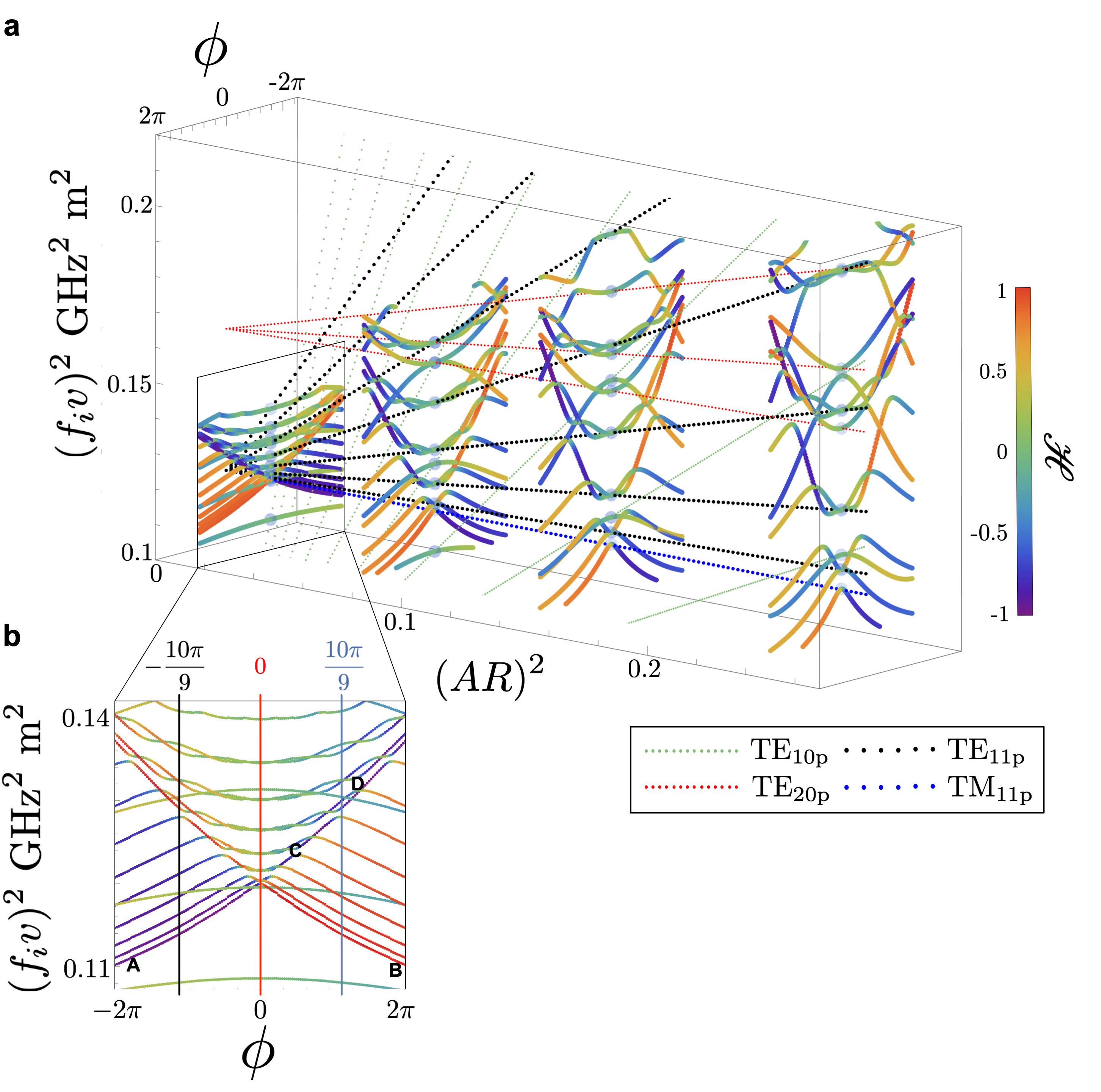}
    \caption{(a) The normalised eigenfrequencies $(f_i v)^2$ of the resonant modes in an equilaterial triangular resonator as a 
    function of $\phi$ for four different $AR$ values, with solution colour corresponding to $\mathscr{H}_i$, and the green, black, red and blue dashed lines 
    corresponding to the TE$_{10p}$, TE$_{11p}$, TE$_{20p}$, \& TM$_{11p}$. (b) The case of $AR=0.133$ $((AR)^2=0.0178)$. Labels A, C and D lie along the $\psi^+_1$ mode, and label B lies on the $\psi^-_1$ mode.}
    \label{fig:norm_plots}
\end{figure}

\section{Origin of Single Modes with Non-Zero Helicity due to Twisting} \label{sec:single_mode_helicity}

\begin{figure*}[t]
    \includegraphics[width=1.7\columnwidth]{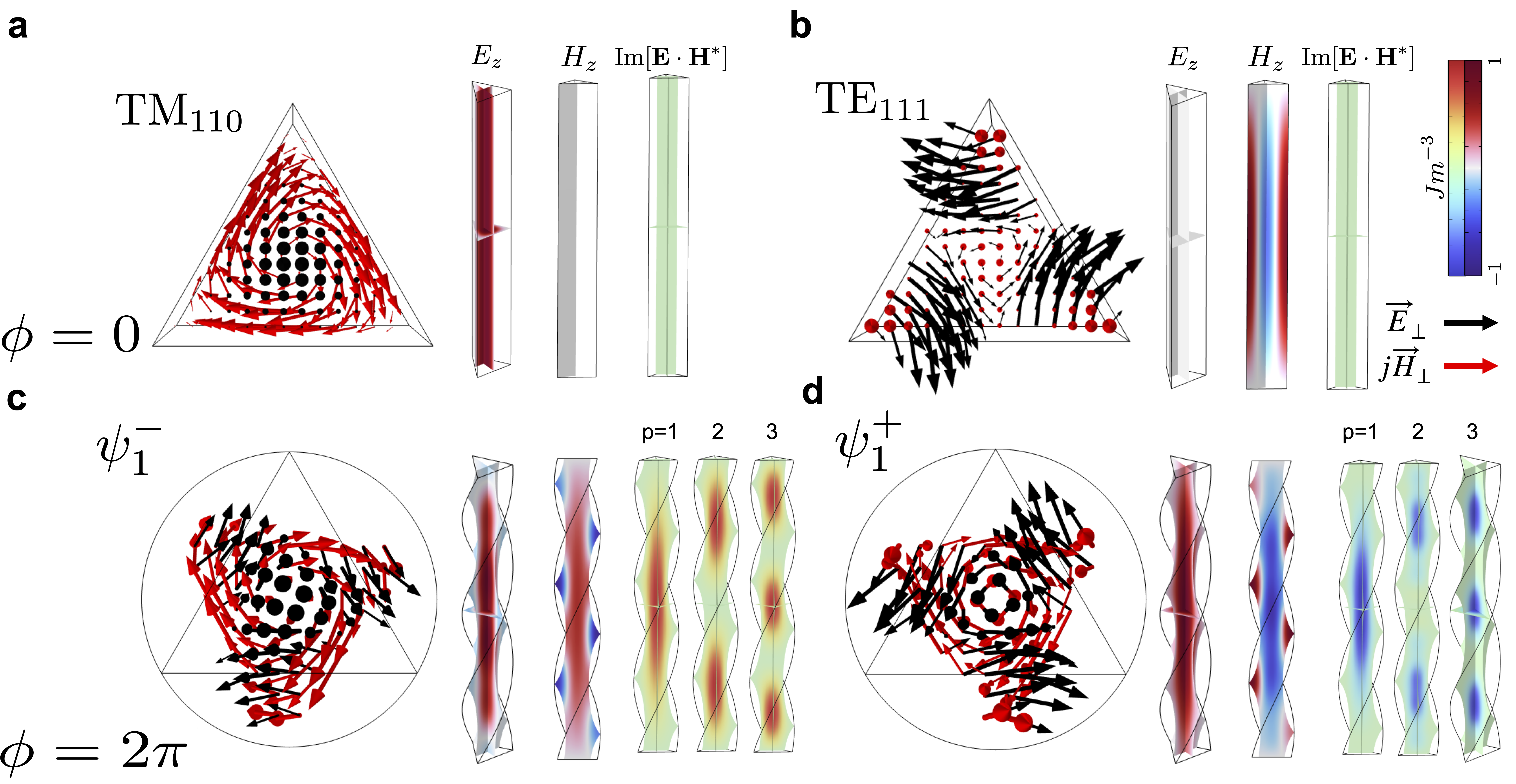}
    \caption{The $\vec{E}_\perp$ (black) \& $j\vec{H}_\perp$ (red) fields and the normalised densities of the axial fields $E_z$, $H_z$ \& $\text{Im}\left[\mathbf{E}_i\cdot\mathbf{H}_i^*\right]$ for the (a) TM$_{110}$ mode, (b) TE$_{111}$ mode, and the twisted ($\phi=2\pi$) (c) $\psi^-$ mode, and (d) $\psi^+$ mode. Note, the mode number $p$ is easily counted by observing the number of maximums in density plot of $|\text{Im}\left[\mathbf{E}_i\cdot\mathbf{H}_i^*\right]|$.
    }
    \label{fig:twisted_modes}
\end{figure*}

\noindent A generic resonator geometry is represented in Fig.~\ref{fig:EH_Relationship}(a) where $v$ is the vertex length of the triangular cross-section, $l$ is the length of the resonator and $\phi$ the angle that the cross-section is rotated over the length of the resonator. The eigenfrequencies $f_i$ of these systems were solved using finite 
element modelling (FEM), for various aspect ratios ($AR=v/l$) 
as a function of $\phi$, with results shown in Fig.~\ref{fig:norm_plots}(a).
Previous studies have examined the modes in un-twisted triangular cross-section prisms of infinite length~\cite{Lopez_Electromagnetic_2016,Analytical_Expressions_Lopez_Moran_2019}, and
have shown that the normalised eigenfrequencies of the form $(f_i v)^2$ are linearly dependent on the square of the aspect ratio $(AR)^2$.
These $\phi=0$ solutions are plotted as the dashed lines in Fig.~\ref{fig:norm_plots}(a) as a function of $(AR)^2$. The colour density of this figure represents $\mathscr{H}_i$ of the resonant modes, 
calculated by evaluating the integrals in~\eqref{eq:helicity} from the FEM solutions.
The frequencies of the resonant modes in the non-twisted resonators with equilateral cross-section are given by: 
\begin{equation}
f_{m,n,p}=\frac{2}{3} c \sqrt{\left(\frac{m}{v}\right)^2+\left(\frac{n}{v}\right)^2+\left(\frac{m n}{v}\right)^2+\left(\frac{p}{l}\right)^2},
\label{eq:fmnp}
\end{equation}
where $c$ is the speed of light, and $m$, $n$, and $p$ are integer mode numbers (here $i\equiv m,n,p$). The mode numbers $m$ and $n$ represent the number of variations in the standing wave pattern in the transverse plane of the cavity, while $p$ corresponds to the number of variations in the longitudinal direction. For a given set of mode numbers $(m, n, p)$, the matching transverse electric (TE) and transverse magnetic (TM) modes are thus degenerate in frequency. The selection rules for the TE modes are $m \geq n \geq 0$, $m \neq 0$, and $p > 0$, while for the TM modes: $m \geq n > 0$ and $p \geq 0$. The straight lines generated by~\eqref{eq:fmnp} on Fig. \ref{fig:norm_plots} intersect the $(f_i v)^2$ vs $\phi$ plots at $\phi=0$, as is expected. 
However, as $\phi$ is increased the majority of the eigenmodes asymmetrically detune in opposite directions from their untwisted frequencies, 
with a corresponding increase in the magnitude of their $\mathscr{H}_i$. For positive $\phi$ we observe negative $\mathscr{H}_i$ modes increasing in frequency whilst positive $\mathscr{H}_i$ modes decrease, and vice-versa for negative $\phi$.

We observe that as $\phi$ increases in magnitude, the near degenerate TM$_{11p}$ and TE$_{11(p+1)}$ modes mix to produce two seperate modes with non-zero $\mathscr{H}_i$, which approaches unity for relatively small angles. We denote modes produced via in-phase mixing of the form TM$_{11p}+$TE$_{11(p+1)}$ as $\psi^+_{p+1}$ and modes produced  via out-of-phase mixing of the form TM$_{11p}-$TE$_{11(p+1)}$ as $\psi^-_{p+1}$. Here given we only investigate the mixing of the $m=n=1$ mode family, we have dropped these subscripts in our $\psi^\pm$ nomenclature. Note that the $\psi^+_{p+1}$ modes will have negative $\mathscr{H}_i$ due to the anti-parallel fields of the untwisted modes, and vice-versa for $\psi^-_{p+1}$. The lowest order $\psi^+_1$ and $\psi^-_1$ modes are labelled A and B, respectively, in the normalised eigenfrequency 
spectra for $(AR)^2=0.0178$ in Fig.~\ref{fig:norm_plots}(b).

The mixing occurs due to the introduction of mirror-asymmetry to the boundary conditions due to non-zero $\phi$. This breaking of spatial symmetry introduces a magneto-electric coupling that mixes the $\vec{\mathbf{E}}_i$ and $\vec{\mathbf{H}}_i$ fields of the TE and TM modes of the untwisted resonator, with examples of the FEM simulated fields shown in Fig.~\ref{fig:twisted_modes}(a) and (b). This mixing effect can be described by the mixing angle $\eta$ that transforms the fields in the resonator according to the 
dual transformation~\cite{Alpeggiani_Electromagnetic_2018,Elektrodynamik2018AxionEA,Planelles_Axion_2021}:
\begin{equation}
    \binom{\vec{\mathbf{E}}_i}{c\mu_0 \vec{\mathbf{H}}_i}=\left(\begin{array}{cc}
        \cos (\eta) & \sin (\eta) \\
        -\sin (\eta) & \cos (\eta)
    \end{array}\right)\binom{\vec{\mathbf{E}}_{\mathbf{0}}}{c\mu_0 \vec{\mathbf{H}}_0}, \label{eq:duality_transform}
\end{equation}
where $\vec{\mathbf{E}}_0$ and $\vec{\mathbf{H}}_0$ are the original vector fields in the untwisted resonator. The duality transformation is real and occurs in the time domain. 

The $\psi^\pm_{p+1}$ modes are shown in Fig.~\ref{fig:twisted_modes}(c) and (d), and form a new orthogonality basis. The normalised field maxima can only be counted when evaluating the field product $\text{Im}\left[\vec{\mathbf{E}}_i\cdot\vec{\mathbf{H}}_i^*\right]$ to determine the mode number $p$. The sign of this field product directly maps to the new orthogonality basis (positive for $\psi^-_{p+1}$, negative for $\psi^+_{p+1}$).

Note that in Fig. \ref{fig:norm_plots}, there also exists modes that do not tune to the same extent as $\phi$ is varied and have relatively low $\mathscr{H}_i$ values. 
These modes are of the $\mathrm{TE}_{10 (p+1)}$ family, which do not have the TM counterpart required to couple together, as dictated by the afformentioned selection rules (i.e. $\mathrm{TM}_{10 p}$ is forbidden) and hence do not mix to produce a larger value of $\mathscr{H}_i$. 

The $\psi^\pm$ modes can be more precisely described by a fractional mixing of the states $|\text{TM}_{11p}\rangle$ and $|\text{TE}_{11(p+1)}\rangle$ given by 
\begin{equation}
    \left|\psi^\pm_{p+1}\right\rangle = \left|\delta\right|\left| \text{TM}_{11p} \right\rangle \pm \left|\beta\right|\left| \text{TE}_{11(p+1)} \right\rangle,
    \label{eq:psipm_state}
\end{equation}
where $\delta$ and $\beta$ are weighting factors that measure the field overlap between the twisted resonant modes and the untwisted resonant modes $\text{TM}_{11p}$ and $\text{TE}_{11(p+1)}$. 
The exact forms of these expressions are given by
\begin{equation}
    \delta = \frac{\int \mathbf{E}(\vec{r})_{\text{TM}_{11p}} \cdot \mathbf{E}(\vec{r})_{\psi^\pm_{(p+1)}} d V}{\int \mathbf{E}(\vec{r})_{\text{TM}_{11p}} \cdot \mathbf{E}(\vec{r})_{\text{TM}_{11p}} \mathrm{dV}},
\end{equation}
and
\begin{equation}
    \beta = \frac{\int \mathbf{H}(\vec{r})_{\text{TE}_{11(p+1)}} \cdot \mathbf{H}(\vec{r})_{\psi^\pm_{(p+1)}} d V}{\int \mathbf{H}(\vec{r})_{\text{TE}_{11(p+1)}} \cdot \mathbf{H}(\vec{r})_{\text{TE}_{11(p+1)}} \mathrm{dV}},
\end{equation}
where $\mathbf{E}(\vec{r})_{\psi^\pm{(p+1)}}$ represents the electric field vector of the twisted $\psi^\pm_{p+1}$ modes, and similar notation applies for the magnetic field vectors. The same notation is used for the untwisted modes.

\begin{figure}[t]
    \centering
    \includegraphics[width=1\columnwidth]{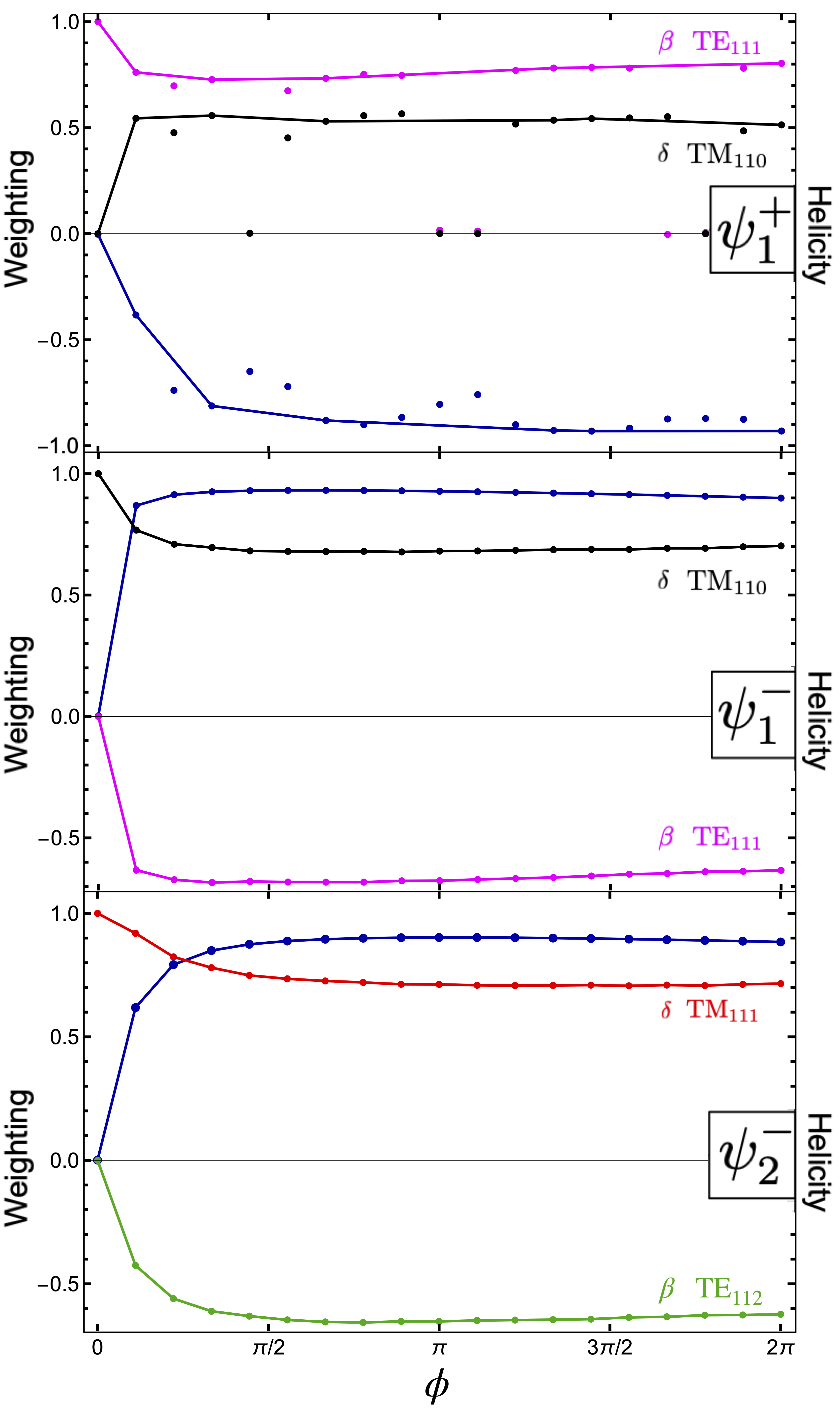}
    \caption{The weighting factors $\delta$ (black) and $\beta$ (pink) of (a) $|\psi_1^+\rangle$ and (b) $|\psi_1^-\rangle$,
    and the corresponding $\mathscr{H}_i$ (blue) as a function of $\phi$ for the equilaterial triangular cross-section resonator.
    Similarly, the $\delta$ (red) and $\beta$ (green) weighting factors for the (c) $\psi_2^-$ mode.}
    \label{fig:Mixing_Plots}
\end{figure}

Fig.~\ref{fig:Mixing_Plots} depicts how the weighting factors of the first three $\psi^\pm_{p+1}$ modes change as a function of $\phi$ and the associated impact on their $\mathscr{H}_i$. As expected, at $\phi=0$, $\beta$ dominates for the $\psi_1^+$ mode as it exists as a TE$_{111}$ mode, and $\delta$ dominates 
for the $\psi_1^-$ mode as it exists as a TM$_{110}$ mode. As $\phi$ is increased, the dominant weighting factor reduces in magnitude 
while its counterpart increases demonstrating how the dominant mode mixes with 
its partner mode, with a corresponding increase in the magnitude of the mode's $\mathscr{H}_i$. 
For the $\psi_1^+$ mode, $\beta$ is positive meaning that the $E_z$ field of the TE$_{111}$ mode 
is being mixed with the $H_z$ of the TM$_{110}$ mode, which in their untwisted forms are anti-parallel, 
therefore the vector product of the two results in a negative $\mathscr{H}_i$. Whereas, for the $\psi_1^-$ mode, $\delta$ and $\beta$ have opposite signs 
meaning that the $E_z$ and $H_z$ fields being mixed are parallel in their untwisted forms, therefore the vector product of the two results in a positive $\mathscr{H}_i$.
The same phenomenon occurs for the higher-order modes (e.g. $\psi^-_2$), where the next pair of near degenerate modes, TM$_{111}$ and TE$_{112}$, mix. 

The mode mixing gets stronger as $\phi$ increases and reaches a maximum at the point at which $\delta$ and $\beta$ are the closest in magnitude, corresponding to the strongest coupling strength, indicated by the red line in Fig. \ref{fig:Mixing_Plots}. 
Note that in Fig.~\ref{fig:Mixing_Plots}(a), the dips in weighting factor and $\mathscr{H}_i$ of the $\psi^+_1$ mode are due to mode interactions with $\psi^-_{p+1}$ modes at certain twist angles, 
which can also be seen at points C and D in Fig.~\ref{fig:norm_plots}(b). As a result of these interactions, the simulated data points deviate from the fitted lines for numerous $\phi$ values and it is difficult to say with certainty at which $\phi$ value a maximum $\mathscr{H}_i$ is achieved.

\subsection{Experimental Validation}
\begin{figure}[t]
    \includegraphics[width=1\columnwidth]{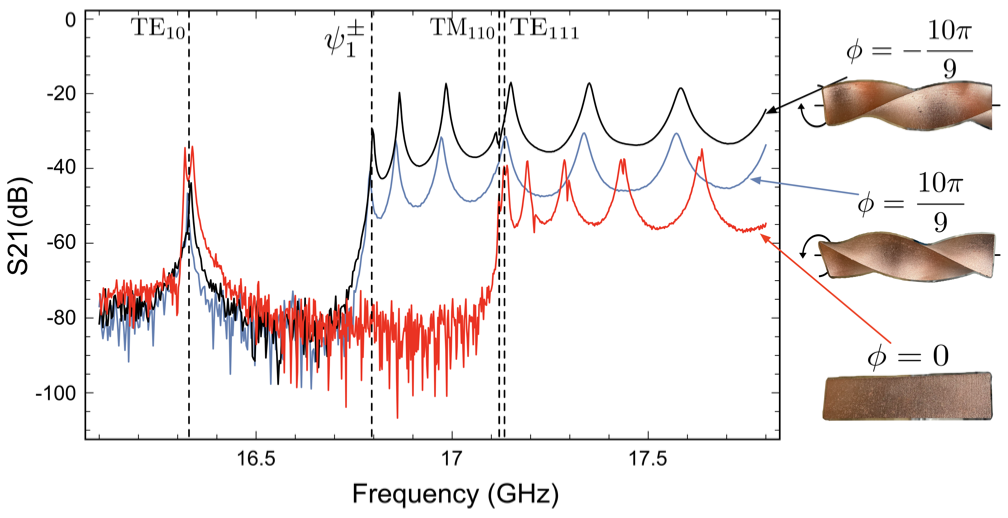}
    \caption{The experimental S$_{21}$ transmission measurements taken from 3D printed copper resonators with dimensions $v=20$ mm and $l=150$ mm at 
    twist angles of $-10\pi/9$ (black), $10\pi/9$ (blue) and $0$ (red), corresponding to the vertical cut lines in Fig.~\ref{fig:norm_plots}(b). 
    The gridlines indicate the corresponding locations 
    of the resonant transmission peaks of the experimental data, specifically the modes TE$_{10}$ for all twist angles, 
    TM$_{110}$ \& TE$_{111}$ for the untwisted resonator and $\psi^\pm$, for the twisted resonators are indicated by dashed lines.}
    \label{fig:copper_plot}
\end{figure}
\noindent To validate the resonant frequency splitting caused by twisting in the triangular resonator, three copper cavity prototypes were 3D-printed using 
selective laser melting. The cavity dimensions were set to $v=20$ mm, $l=150$ mm, and therefore $(AR)^2=0.0177$, with one resonator having a 
right-handed twist ($\phi=10\pi/9$), another a left-handed twist ($\phi=-10\pi/9$), and a third untwisted (\(\phi=0\)). These 
dimensions were chosen to ensure the prototypes were easily printable. 

Transmission ($S_{21}$) measurements were obtained  
through vector network analysis across two magnetic field antenna placed at the cavity end 
faces, with results shown in Fig.~\ref{fig:copper_plot}. As predicted the $\psi^\pm_{p+1}$ modes in the twisted resonators 
are shifted down in frequency compared to the origial TE$_{110}$ and TM$_{111}$ modes in the untwisted resonator. 
By comparison, the lower order TE$_{10}$ mode, which, as mentioned previously, does not undergo any mixing effect due to twisting, 
has the same resonant frequency in both the untwisted and twisted resonators. 
In addition, we observe that the frequency shift due to twisting is equal in magnitude for both the left and right handed resonators, 
confirming that the tuning effect is symmetric about $\phi=0$, as predicted in Fig.~\ref{fig:norm_plots}(b), where the twist 
angles of the experimental cavities are shown by the grid lines. The observed frequency shift of the lowest frequency $\psi$ mode between the $\phi=10\pi/9$ and $\phi=0$ resonators is measured to be $-335$ MHz, whilst the frequency shift predicted for $AR=0.0177,~v=20~$mm from Fig. \ref{fig:norm_plots}(b) is $\sim-379$ MHz {--} in reasonable agreement given the printed dimensions may not exactly match those that were simulated.

\section{Generation of Helicity due to Material Parameter $\kappa$}\label{sec:helicity_kappa}

\noindent Electromagnetic helicity commonly appears when photons interact with chiral molecules or materials \cite{Cohen2010,Tang2011,Hendry:2010ug}. 
The chiral strength of these materials is quantified by the parameter $\kappa$, which modifies Maxwell's equations by introducing an imaginary 
coupling term between the electric and magnetic fields. This magnetoelectric coupling results from $\kappa$ breaking spatial symmetry, 
leading to a mixing of the perpendicular electric \textbf{E} and magnetic \textbf{H} fields in the material and therefore a non-zero $\text{Im}\left[\mathbf{E}\cdot\mathbf{H}^*\right]$ product.
From the form of~\eqref{eq:local_helicity}, this will 
result in a non-zero, $h$, akin to the effect of twisting the resonator.
The effect of $\kappa$ is in contrast to a real, in-phase coupling between electric and magnetic fields, which is represented by the non-reciprocity 
parameter $\chi$, which rather than generating helical radiation, generates nonreciprocity. 
Specifically, $\chi$ 
influences the phase of the propagating electric field without altering its polarisation, resulting in a time-dependent non-orthogonality with the magnetic field. On the other hand, 
$\kappa$ affects the polarisation of the propagating field without changing its phase, leading to a spatial non-orthogonality with the magnetic field~\cite{Electromagnetic_Waves_Textbook_1994}.
A positive $\kappa$ indicates left-handed polarisation rotation in the direction of propagation, whilst a negative $\kappa$ indicates right-handed rotation. 
A material with non-zero $\kappa$ and $\xi=0$ is referred to as a Pasteur or chiral medium, whilst a material with non-zero $\chi$ but $\kappa=0$ is 
referred to as a Tellegen material. The general case with both parameters non-zero is referred to as {``}bi-isotropic{''}, which will have the following constitutive equations inside the medium:
\begin{equation}
    \begin{aligned}
        \mathbf{D} &= \epsilon \mathbf{E} + \xi \mathbf{H}, \\
        \mathbf{B} &= \zeta \mathbf{E} + \mu \mathbf{H} \label{eq:CE_kappa},
    \end{aligned}
\end{equation} 
where $\epsilon$ is the dielectric permittivity of the medium, $\mu$ is the magnetic permeability of the medium, $\mathbf{D}$ is the electric displacement field 
and $\mathbf{H}$ is the inductive magnetic field.  
The coupling between the electric and magnetic fields is characterised by the parameters:
\begin{align}
    \xi&=(\chi-j \kappa) \sqrt{\mu_0 \epsilon_0},\\
    \zeta&=(\chi+j \kappa) \sqrt{\mu_0 \epsilon_0},
    \label{eq:coupling_parameters}
\end{align}
with $\epsilon_0$ and $\mu_0$ are the permittivity and permeability of free space, respectively. These constitutive equations are presented in the frequency domain.

\section{Perturbation Theory}\label{sec:freq_shift}

\noindent It should be noted that in the case of a mirror-symmetric resonator, a frequency shift will be produced by the introduction of a chiral material if the 
perturbed $\mathbf{E}$ and $\mathbf{H}$ fields in~\eqref{eq:perturb_freq_shift} are significantly altered from their original states. This will only be true when the companion field generated by the electro-magnetic 
coupling introduced by the media; i.e.  $j\kappa\sqrt{\mu_0 \epsilon_0}\text{H}_z$ originating from an applied resonant E$_z$, is itself resonant. This is the case for the 
TM$_{11p}$ and TE$_{11(p+1)}$ modes in the cavities being discussed due to the equilateral cross section and hence their near-degenerate frequencies. 

From perturbation theory, for a cavity resonator with bi-isotropic inclusions, there is a known relationship between the pertubed value $\kappa_1$ and a shift in resonant frequency $\Delta \omega=\omega_1-\omega_0$, where $\omega_1$ is the new resonant frequency and $\omega_0$ is the unperturbed cavity frequency. 
Many studies suggest this relationship as a means of measuring the value of $\kappa$ ~\cite{Electromagnetic_Waves_Textbook_1994,Tretyakov_waveguide_resonator_1995,kong1990electromagnetic}. 
This perturbation has previously only been considered for the inclusion of a small bi-isotropic sphere into a resonator resulting in small deviations around the isotropic properties $\kappa_0=0$ and $\xi_0=0$~\cite{Electromagnetic_Waves_Textbook_1994}. 
Here, we derive the relationship between $\Delta\omega$ and $\kappa_1$ using perturbation around an arbitrary $\kappa_0$ in terms of the mode helicity. 

Consider a resonator filled with some bi-anistropic material that has some small change in the chiral material parameter $\Delta\kappa=\kappa_1-\kappa_0$. 
The unperturbed electric and magnetic fields $\mathbf{E}_0$, $\mathbf{H}_0$ and the perturbed fields $\mathbf{E}_1$, $\mathbf{H}_1$ inside the resonator all satisfy Maxwell{'}s equations (with no source terms), i.e. 
\begin{equation}
	\begin{aligned}
	\nabla\times\mathbf{E}&=-j\omega\mathbf{B}~\text{and}~\nabla\times\mathbf{E}^*=j\omega\mathbf{B}^*, \\
	\nabla\times\mathbf{H}&=j\omega\mathbf{D}~\text{and}~\nabla\times\mathbf{H}^*=-j\omega\mathbf{D}^*.
	\end{aligned}
	\label{eq:ME}
\end{equation}

The analysis begins with considering the two electromagnetic processes $(\mathbf{E}_0, \mathbf{H}_0, \mathbf{D}_0, \mathbf{B}_0)$ and $(\mathbf{E}_1, \mathbf{H}_1, \mathbf{D}_1, \mathbf{B}_1)$ at different oscillation frequencies $\omega_0$ and $\omega_1$, and considering the following combinations of vector fields,
\begin{equation}
    \begin{aligned}
  & \nabla\cdot (\mathbf{E}_0^*\times \mathbf{H}_1+\mathbf{E}_1 \times \mathbf{H}_0^*)= \mathbf{H}_1\cdot\nabla\times \mathbf{E}_0^*\\
  &-\mathbf{E}_0^*\cdot\nabla\times \mathbf{H}_1+ \mathbf{H}_0^* \cdot\nabla\times \mathbf{E}_1 - \mathbf{E}_1\cdot\nabla\times \mathbf{H}_0^* 
    \label{eq:S}
    \end{aligned}
\end{equation}
Combining (\ref{eq:CE_kappa}) and (\ref{eq:ME}) with (\ref{eq:S}) we obtain,
\begin{equation}
    \begin{aligned}
  &\nabla\cdot (\mathbf{E}_0^*\times \mathbf{H}_1+\mathbf{E}_1 \times \mathbf{H}_0^*)=j\omega_0\mu \mathbf{H}_0^*\cdot\mathbf{H}_1+j\omega_0\zeta_0^* \mathbf{E}_0^*\cdot\mathbf{H}_1\\
   &-j\omega_1\epsilon \mathbf{E}_1\cdot\mathbf{E}_0^*-j\omega_1\xi_1 \mathbf{H}_1\cdot\mathbf{E}_0^*- j\omega_1\mu \mathbf{H}_1\cdot\mathbf{H}_0^*\\
   &-j\omega_1\zeta_1 \mathbf{E}_1\cdot\mathbf{H}_0^*+j\omega_0\epsilon \mathbf{E}_0^*\cdot\mathbf{E}_1+j\omega_0\xi_0^* \mathbf{H}_0^*\cdot\mathbf{E}_1
    \label{eq:S2}
    \end{aligned}
\end{equation}
where $\zeta_0$ and $\xi_0$ are the unperturbed bi-isotropic parameters of the system (i.e. with $\kappa_0$) and $\zeta_1$ and $\xi_1$ the perturbed parameters (i.e. with $\kappa_1$). 

Next we set the non-reciprocity parameters, $\chi_0,\chi_1=0$, in order to considering a Pasteur material with non zero $\kappa$, then we apply perturbation theory where we assume  $\mathbf{H}_1= \mathbf{H}_0$ and $\mathbf{E}_1= \mathbf{E}_0$. Given the complex Poynting vector, $\mathbf{S}_0=1/2(\mathbf{E}_0\times\mathbf{H}^*_0)$, and integrating over the volume, (\ref{eq:S2}) becomes. 
\begin{equation}
    \begin{aligned}
  4\int\text{Re}&\left(\nabla\cdot \mathbf{S}_0\right)dV=\int-j(\omega_1-\omega_0)(\mu |\mathbf{H}_0|^2+\epsilon |\mathbf{E}_0|^2)\\
   &+j\sqrt{\epsilon_0\mu_0}(\omega_1\kappa_1-\omega_0\kappa_0)2\text{Im}[\mathbf{E}_0\cdot\mathbf{H}_0^*]~dV
    \label{eq:S3}
    \end{aligned}
\end{equation}
In the ideal lossless scenario, the left hand side of (\ref{eq:S3}) is zero, in this case, to first order in perturbations we may show,
\begin{equation}
\delta\omega=\frac{\omega_0\delta\kappa+\delta\omega\kappa_0}{2\mu_r\epsilon_r}\mathscr{H}_0
\end{equation}
and with some rearanging
\begin{equation}
\frac{\delta\omega}{\omega_0}=\frac{\delta\kappa}{\kappa_0}\left(\frac{\frac{\kappa_0\mathscr{H}_0}{2\mu_r\epsilon_r}}{1-\frac{\kappa_0\mathscr{H}_0}{2\mu_r\epsilon_r}}\right),
\label{eq:perturb_freq_shift}
\end{equation}
where the helicity of the unperturbed mode, $\mathscr{H}_0$, is given by (\ref{eq:helicity}) and $\epsilon_r$ and $\mu_r$ are the relative permittivity and permeability of the materials respectively.
In the case $\kappa_0\ll1$ then (\ref{eq:perturb_freq_shift}) becomes,
\begin{equation}
\left(\frac{\delta\omega}{\omega_0}\right)_{\kappa_0\ll1}=\frac{\delta\kappa\mathscr{H}_0}{2\mu_r\epsilon_r}.
\label{eq:perturb_freq_shift0}
\end{equation}
The $\psi_1^{-}$ mode has $\mathscr{H} > 0$ for $\phi > 0$, and since this corresponds to the same handedness as radiation with $\kappa < 0$,  \eqref{eq:perturb_freq_shift0} predicts that the $\psi_1^{-}$ mode will shift to a lower frequency, i.e., $\frac{\delta\omega}{\omega_0}$. This is consistent with the FEM-simulated eigenfrequencies shown in Fig.~\ref{fig:norm_plots}(b). 

Rearranging (\ref{eq:perturb_freq_shift}) we may also represent the helicity in term of the frequency perturbation as,
 \begin{equation}
 \mathscr{H}_0=\frac{\frac{2\mu_r\epsilon_r}{\omega_0}\frac{\delta\omega}{\delta\kappa}}{1+\frac{\kappa_0}{\omega_0}\frac{\delta\omega}{\delta\kappa}}
 \label{incrHel}
 \end{equation}
 Thus the helicity may be calculated either from the volume integral of the fields as indicated in (\ref{eq:helicity}) or through an incremental frequency rule as indicated in (\ref{incrHel}). This is very similar to the way electromagnetic filling factors of resonant modes may be numerically calculated, that is via volume integrals or an incremental frequency rule as indicated in \cite{Krupka_Whispering_Gallery_1999}.

\begin{figure}[t]
    \includegraphics[width=0.98 \columnwidth]{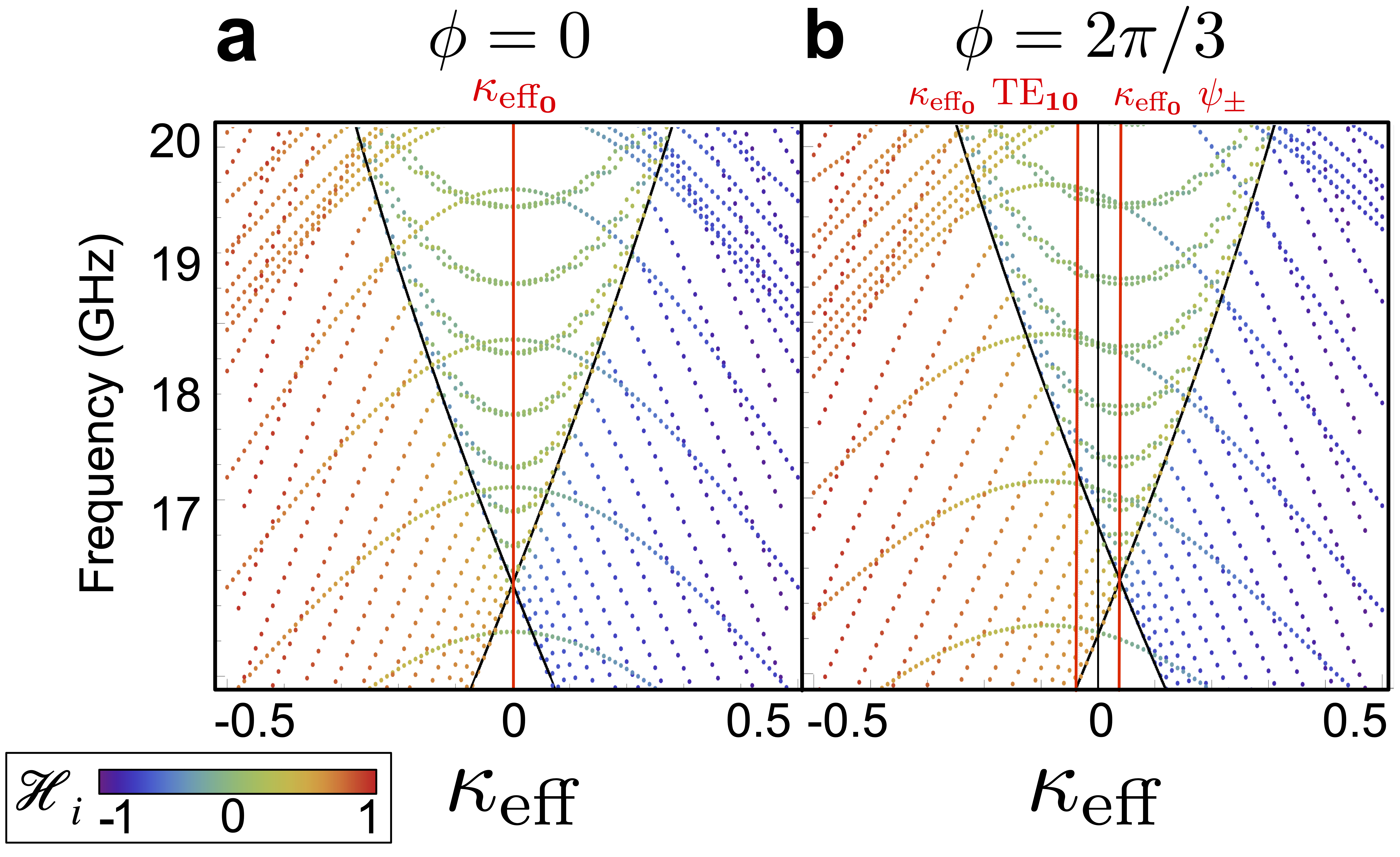}
    \caption{Eigenfrequency solutions for the $v=20$ mm and $l=150$ mm cavity as a function of the material parameter 
    $\kappa_\text{eff}$ for twist angles of (a) $0$ and (b) $\frac{2\pi}{3}$ where the colour is $\mathscr{H}_i$. The red grid lines mark the 
    $\kappa_\text{eff}$ values at which there is zero $\mathscr{H}_i$.}
    \label{fig:kappa_axion_field_2}
\end{figure}

From the form of \eqref{eq:perturb_freq_shift0}, we expect a linear relationship between $\kappa\simeq \kappa_\text{eff}$ and $\Delta\omega/\omega$ if we assume $\mathscr{H}_i$ is a constant.  
We can evaluate this relationship by solving an electromagnetic FEM of an untwisted triangular resonator with modified Maxwell's equations 
according to (\ref{eq:CE_kappa}) and sweeping the chirality parameter, $\kappa_\text{eff}$. The resulting eigenfrequencies and $\mathscr{H}_i$ 
calculated from the eigenvectors are shown in Fig. \ref{fig:kappa_axion_field_2}(a) for a $v=20$ mm, $l=150$ mm and $\phi=0$ triangular cross-section 
cavity. We observe that for a given eigenmode, once the $\mathscr{H}_i$ saturates to a constant (black lines), we indeed observe a linear relationship between 
frequency and $\kappa_\text{eff}$ proportional to the $\mathscr{H}_i$ value. Importantly, (\ref{eq:perturb_freq_shift0}) demonstrates the relationship between frequency shift and electromagnetic helicity, where the rate of change of frequency with respect to material chirality ($\delta\omega/\delta\kappa$) is determined by mode helicity at a given $\kappa_0$. 

Furthermore, it is interesting to investigate the above situation in a twisted cavity, as shown in Fig. \ref{fig:kappa_axion_field_2}(b) for 
$\phi=2\pi/3$. We note that the symmetry around $\kappa_\text{eff}=0$ is broken in this case with a new symmetry axis (shown in red) existing at 
offset $\kappa_\text{eff}$ values, which are different for the $\psi^\pm$ modes and the TE$_{10}$ modes. This result points to a 
mode-family-dependent equivalency between twisting and $\kappa$ where the two effects have the potential to cancel one another at certain values. Importantly, 
the result of Fig. \ref{fig:kappa_axion_field_2}(b) is that two materials with the same magnitude $\kappa_\text{eff}$ but different signs 
(i.e. a left- and right-handed form of the same substance) will produce distinguishable spectra in a twisted cavity, whilst they will produce 
identical spectra in an untwisted cavity.

\section{Equivalence between $\kappa_\text{eff}$ and $\phi$}

\begin{figure}[t]
    \centering
    \includegraphics[width=1\columnwidth]{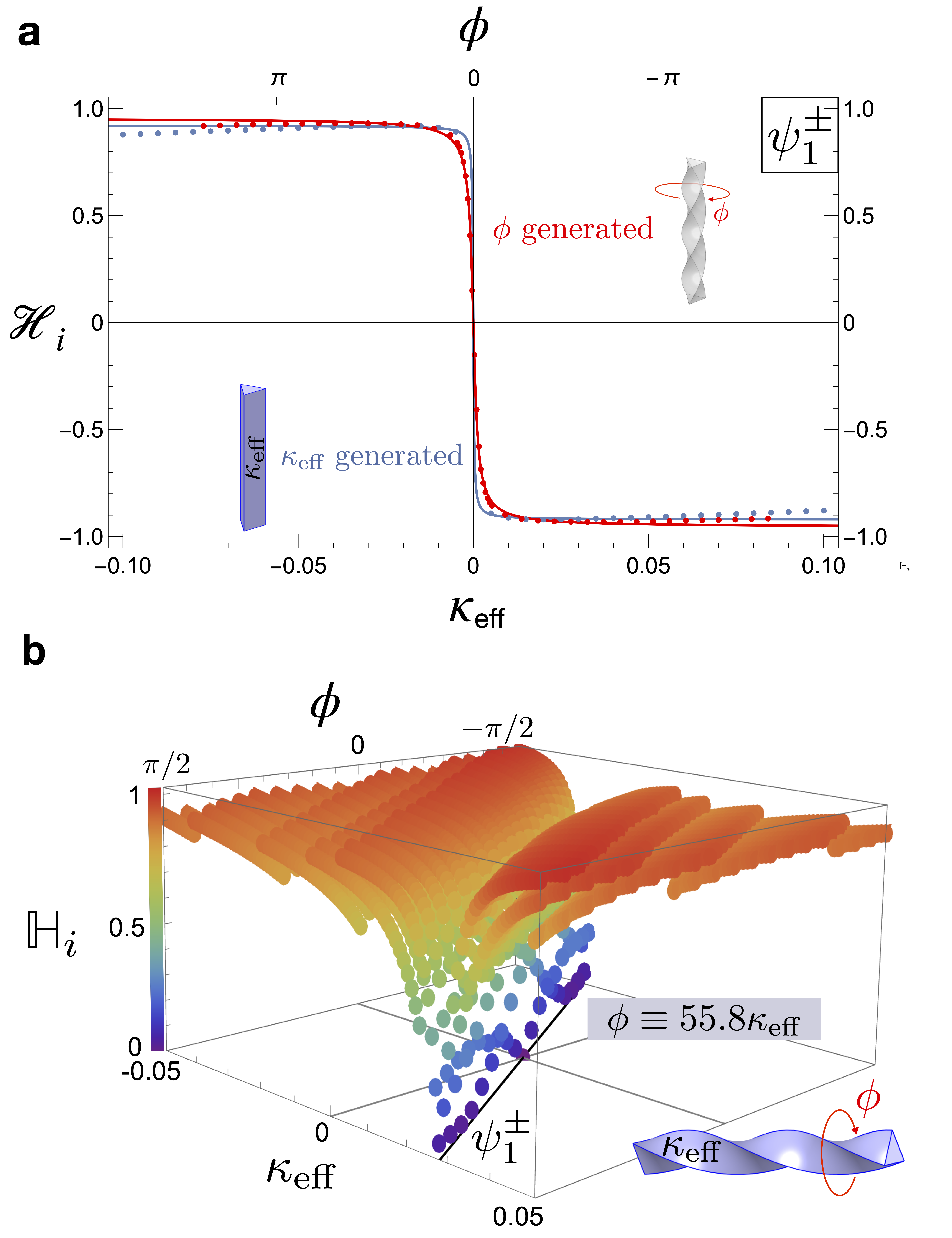}
    \caption{For the equilateral triangular cross-section resonator with dimensions of $v=20$ mm and $l=150$ mm: (a) $\mathscr{H}_i$ of the $\psi_1^\pm$ mode as a 
    function of introduced $\phi$ (red) and $\kappa_\text{eff}$ (blue) to the system. (b) FEM simulated $\mathbb{H}_{ i}$ values as both $\phi$ 
    and $\kappa_\text{eff}$ are varied demonstrating null points where the helicity induced by one effect is cancelled by the other.} 
    \label{fig:kappa_axion_field}
\end{figure}

\noindent Given that $\mathscr{H}_i$ values resulting from twisting the triangular cross-section resonators are purely 
real (i.e. $\vec{\mathbf{E}}\cdot\vec{\mathbf{H}}^*$ is purely imaginary), we 
can safely assume that the process of twisting can be alternatively described by the introduction of some $\kappa$ value for the 
propagation media whilst setting $\chi=0$. We can demonstrate the explicit equivalence between a $\phi$ twist and filling the 
cavity with some isotropic chiral media by firstly simulating these two distinct cases. From the solved eigenvectors we can plot $\mathscr{H}_i$ as a function of $\kappa_\text{eff}$ and twist angle $\phi$ as shown in Fig.~\ref{fig:kappa_axion_field}(a). We see that these two results agree when an appropriate, phenomenological scaling is applied to the $x$-axis. 

Furthermore, the true equivalence of the two mechanisms can be demonstrated through nullifying the $\mathscr{H}_i$ generated by twisting 
with an opposite acting $\kappa_\text{eff}$. This effectively represents how a uniform chiral media surrounded by conducting boundary conditions can be twisted to produce linearly polarised, un-coupled TE and TM modes and corresponds to the situation represented by the red  $\kappa_{\text{eff}_0}$ line in Fig. \ref{fig:kappa_axion_field_2}(b) for $\phi=2\pi/3$. To accurately locate these null-points we must evaluate the integral of $|h|$ over the volume, which is written as:
\begin{equation}
    \mathbb{H}_{ i}=\int |h_p| dV. \label{eq:abs_H}
\end{equation}
The reason for this is that artificial null points can arise due to mode interferences whereby $\mathscr{H}_i$ over 
one portion of the cavity volume will cancel $\mathscr{H}_i$ in the remaining volume and it is not strictly $0$ at every point. 

These {``}de-hybridisation{''} points are tracked in Fig.~\ref{fig:kappa_axion_field}(b). As expected, 
the $\kappa_\text{eff}$ required to nullify $\mathbb{H}_{ i}$ introduced by twisting is linearly proportional to $\phi$, where 
a left handed $\kappa_\text{eff}$ cancels out a right-handed $\mathscr{H}_i$ and vice versa. Given that a 
right-handed $\mathscr{H}_i$ corresponds to a negative $\kappa_\text{eff}$ but a positive $\phi$, 
a positive $\kappa_\text{eff}$ is required to cancel a positive $\phi$. For the specific cavity dimensions used in the FEM
simulation $(v=20$ mm, $l=150$ mm), $\phi\approx 55.8\kappa_\text{eff}$ for the or the $\psi^-_1$ mode. Whilst the relationship between $\kappa$ and $\delta\omega$ has the analytical form of (\ref{eq:perturb_freq_shift0}), the conversion between $\phi$ and $\kappa$ will be mode dependent and cavity geometry dependent, hence only phenomenological equivalencies can be derived.

\begin{figure}[t]
    \includegraphics[width=\columnwidth]{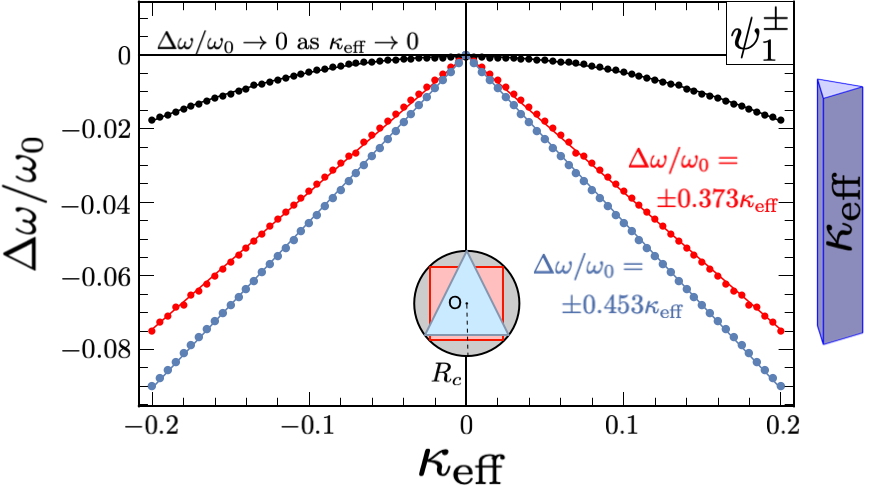}
    \caption{The FEM simulated shift in eigenfrequencies for the $\psi_1^\pm$ resonant modes 
    in the equilaterial triangular resonator (blue) of dimensions $v=20$ mm and $l=150$ mm as a function of $\kappa_\text{eff}$.
    For comparison, the eigenfrequencies for the square (red) and circle (black) cross-sections, which are inscribed 
    within the triangle's circumcircle, are also included.}
    \label{fig:Triangular_Cross_Section}
\end{figure}

As indicated by \eqref{eq:perturb_freq_shift0}, we expect a linear relationship between $\kappa\simeq \kappa_\text{eff}$ if $\mathscr{H}_i$ is constant. It can be seen from Fig.~\ref{fig:kappa_axion_field}(a) that for the chosen cavity dimensions $\mathscr{H}_i$ of the $\psi^\pm_1$ modes saturates out for relatively small $\kappa_\text{eff}$ values at $|\mathscr{H}_i|=0.935$. In this region where $\mathscr{H}_i$ can be assumed constant, we would therefore expect the fractional frequency shift to be linearly dependent on $\kappa$ with a coefficient of $\mathscr{H}_i/2$. Indeed, when taking the eigenfrequency solutions of the FEM in which $\kappa_\text{eff}$ is swept, we can derive the empirical relationship $\Delta\omega/\omega=0.453\kappa_\text{eff}$ for the $\psi^-_1$, which is shown in blue in Fig.~\ref{fig:Triangular_Cross_Section}. 

Since the mode-mixing effect that causes a frequency shift is now proven to be equivalent to the introduction of a twist angle $\phi$ to the metallic boundary conditions, 
we can explore how the cavity cross section plays a role. For a circular cross-section resonator, as a twist is introduced, there will be no impact on the boundary conditions and hence we anticipate no first-order increase in $\mathscr{H}$ and $\Delta\omega/\omega=0$. It therefore stands to reason that for regular polygons of the dihedral group with the same circumcircle, the lowest symmetry order cross-section (i.e. the triangle) will generate the maximal possible frequency shift per change in $\phi$ or $\kappa_\text{eff}$. As a further proof of equivalency between $\phi$ and $\kappa_\text{eff}$, we investigate how changing $\kappa_\text{eff}$ as a material parameter 
over the volume of different geometry cavities affects the resonant frequency of the $\psi^\pm_1$ modes, as shown in Fig.~\ref{fig:Triangular_Cross_Section}. 
Whilst we don't observe an absence of $\Delta\omega/\omega$ for the circular cross-section cavity as we would expect for varying $\phi$, we do indeed observe a minimal amount compared to the maximal for the triangular case. This residual quadratic tuning is a result of higher-order effects that are not taken into consideration in the perturbation theory presented above (which is only to first-order). It is interesting to note that this tuning mechanism appears very similar to the frequency behaviour of the TE$_{01}$ mode family (see Fig.  \ref{fig:norm_plots} and \ref{fig:kappa_axion_field_2}), which would also not be expected to tune given there's no degenerate TM counterpart them, in that it is quadratic and smaller in magnitude compared to the other modes. 


Using the FEM simulations of Fig.~\ref{fig:norm_plots} we can estimate that a $\sim$20 GHz cavity 
with a twist angle of $\phi=2\pi$ will result in $\Delta\omega=-500$ MHz 
for the $\psi^-_1$ mode. Therefore $\Delta\omega/\omega=-0.025$ and such a
cavity would have an equivalent chirality material parameter of $\kappa_\text{eff}=-0.055$. 
For comparison,
Cr$_2$O$_3$, the crystalline material that exhibits the strongest observed magneto-electric effect, 
only has a non-reciprocity parameter of $\chi=3.11\cdot 10^{-4}$~\cite{Hehl_Relativistic_2008} (see appendix~\ref{section:chirality_parameter}).
Note that $\chi$ and $\kappa$ are quadrature terms in the magneto-electric vector and hence it is reasonable to compare their magnitudes. 
The materials with the 
largest $\kappa$ values (ranging from $3.9\cdot 10^{-5}$ to $0.34$), at microwave frequencies, are man-made materials (such as metal helices in epoxy) and are very lossy~\cite{Electromagnetic_Waves_Textbook_1994}, whereas
we generate a modest $\kappa$ value limited only by conductivity losses. 

\section{Generation of Helicity due to Axion Field $\theta$}

\noindent Another method of generating helical (chiral) radiation is to introduce the axion field $\theta$ to Maxwell's equations. 
The constitutive relation for electromagnetism will then be given in the time domain by~\cite{Elektrodynamik2018AxionEA}: 
\begin{equation}
    \begin{aligned}
    \vec{\mathbf{D}} & = \epsilon \vec{\mathbf{E}} - \frac{\theta \alpha}{\pi} c\mu\vec{\mathbf{H}}, \\
    c\vec{\mathbf{B}} & = \frac{\theta \mu \alpha}{\pi} \vec{\mathbf{E}}+\mu\vec{\mathbf{H}},
    \end{aligned} \label{eq:CE_axion}
\end{equation}
where $\alpha=\frac{e^2}{\hbar c}$ is the fine structure constant.
The axion is calculated to couple to photons through the axion-electromagnetic chiral anomaly, whether in condensed matter physics or in QCD. 
The chiral anomaly is described 
by the following interaction terms, added to the Lagrangian of the photonic degrees of freedom of the standard model~\cite{Sikivie2021}:
\begin{equation}
    \mathcal{L}_{a \gamma \gamma}=g_{a \gamma \gamma} a \mu_0 \vec{\mathbf{E}} \cdot \vec{\mathbf{H}}.
    \label{eq:AxInt}
\end{equation}
Here, the axion modification to the equations of
motion are given by the product of the pseudoscalar, $a$, and the photon-axion coupling term $g_{a \gamma \gamma}$ multiplied by $\mathbf{E} \cdot \mathbf{H}$, 
which may look familiar as the numerator of $\mathscr{H}_i$~\eqref{eq:helicity}, multiplied by $j$. 

The square of the magnitude of $\mathscr{H}_i$, $|\mathscr{H}_i|^2$, is equivalent to the form 
factor that couples low-mass axions to photons as an amplitude modulation of a single mode 
through a parametric interaction. This axion-photon coupling due to the twisting of the resonator
is explained in more detail in ~\cite{Bourhill_twisted_anyon_cavity_2023}.

The introduction of the axion field $\theta$ as a modification to Maxwell's equations couples the electric and magnetic fields i.e. magneto-electric coupling, 
mixing them together. This can be 
thought of as a ``rotation'' of the electric and magnetic fields into each other, 
where the rotation angle is the mixing angle $\eta$. The axion field $\theta$ can be related to 
$\eta$ by~\cite{Planelles_Axion_2021,Elektrodynamik2018AxionEA}:
\begin{equation}
    \eta = \tan^{-1}(\frac{\alpha\theta \mu c}{\pi}). \label{eq:mixing_angle}
\end{equation}
This mixing effect is precisely the effect we describe in Section~\ref{sec:single_mode_helicity} due to the twisting of the resonator, 
or in section~\ref{sec:helicity_kappa} by introducing some chirality material parameter $\kappa_\text{eff}$.

We can establish a relationship between the angle $\theta$ and $\vec{\mathbf{E}}^\prime\cdot\vec{\mathbf{H}}^\prime$, where the prime indicates the axion modified fields,
by starting with the inverse of the previously discussed duality transformation~\eqref{eq:duality_transform}:
\begin{equation}
    \left(\begin{matrix}
        \vec{\mathbf{E_0}} \\
        c\mu_0\vec{\mathbf{H_0}}
    \end{matrix}\right)
    =
    \left(\begin{matrix}
        -c\mu_0\sin (\eta) \vec{\mathbf{H}}^\prime+\cos (\eta) \vec{\mathbf{E}}^{\prime} \\
        c\mu_0\cos (\eta) \vec{\mathbf{H}}^\prime+\sin (\eta) \vec{\mathbf{E}}^{\prime}
    \end{matrix}\right).
\end{equation}
Given that the electric and magnetic fields are orthogonal in the untwisted (or $\kappa_\text{eff}=0$ or $\theta=0$) case, their dot product must be zero. 
\begin{equation}
    \begin{aligned}
    0=&\vec{\mathbf{E}}_0 \cdot c\mu_0 \vec{\mathbf{H}}_0 \\
    =&2 c\mu_0 \vec{\mathbf{E}}^{\prime} \cdot \vec{\mathbf{H}}^{\prime} \cos 2 \eta+\sin 2 \eta\left(\vec{\mathbf{E}}^\prime\cdot\vec{\mathbf{E}}^\prime-c^2\mu_0^2 \vec{\mathbf{H}}^\prime\cdot\vec{\mathbf{H}}^\prime\right). \\
    \end{aligned}
\end{equation}
By rearranging to solve for $\eta$, we obtain:
\begin{equation}
    \eta=\frac{1}{2}\tan^{-1}\left(\frac{2c\mu_0 \vec{\mathbf{E}}^\prime\cdot\vec{\mathbf{H}}^\prime}{c^2\mu_0^2 \vec{\mathbf{H}}^{\prime}\cdot\vec{\mathbf{H}}^\prime - \vec{\mathbf{E}}^\prime\cdot\vec{\mathbf{E}}^\prime}\right). \label{eq:eta_h}
\end{equation}
Relating this to the mixing angle~\eqref{eq:mixing_angle}, we derive the relationship between $\vec{\mathbf{E}}_0^\prime\cdot\vec{\mathbf{H}}_0^\prime$ and $\theta$:
\begin{equation}
    \theta = \frac{\pi}{\alpha \mu_0 c}\tan\left(\frac{1}{2}\tan^{-1}\left(\frac{2 c\mu_0 \vec{\mathbf{E}}^\prime\cdot\vec{\mathbf{H}}^\prime}{c^2\mu_0^2 \vec{\mathbf{H}}^{\prime}\cdot\vec{\mathbf{H}}^\prime - \vec{\mathbf{E}}^\prime\cdot\vec{\mathbf{E}}^\prime}\right)\right). \label{eq:axion_EB}
\end{equation}
By the form of~\eqref{eq:local_helicity}, $\vec{\mathbf{E}}^\prime\cdot\vec{\mathbf{H}}^\prime$ in this expression is proportional to $h$. 
This expression can be evaluated at each point within the resonator to qualitatively describe the axion field. For a quantitative measure of axion field strength, 
the integral of~\eqref{eq:axion_EB} can be taken over the volume of the resonator, giving the effective axion field strength $\theta_{\text{eff}}$. 

\begin{figure}[t]
    \includegraphics[width=1\columnwidth]{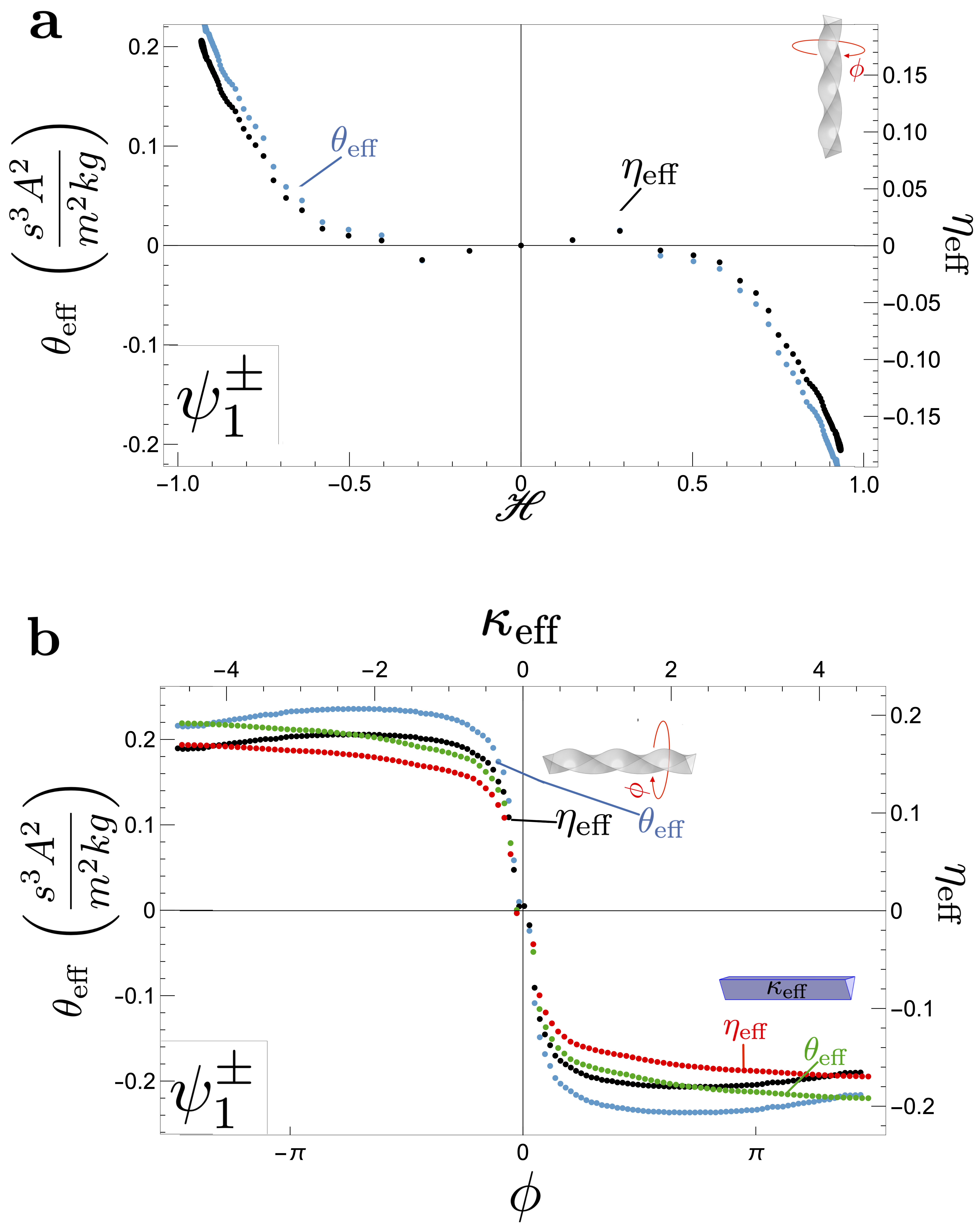}
    \caption{
        (a) The derived results ($\theta_{\text{eff}}$ and $\eta_{\text{eff}}$) are shown in relation to $\mathscr{H}$ for the
        $\psi_1^\pm$ resonant modes derived from the 
        FEM-simulated twisting of an equilaterial triangular resonator with dimensions $v=20$ mm and $l=150$ mm. 
        These derived results ($\theta_{\text{eff}}$ and $\eta_{\text{eff}}$) are shown in (b) as a function of twist 
        angle $\phi$ by the blue and black lines, respectively, along with the derived $\theta_{\text{eff}}$ and $\eta_\text{eff}$
        due to the introduction of the chirality material parameter $\kappa_{\text{eff}}$ over the volume of the resonator, as 
        shown by the green and red lines, respectively.  }
    \label{fig:phi_axion_2}
\end{figure}

Using the small angle approximation ($\tan(x) \approx \arctan(x) \approx 1$), we can relate
equations~\eqref{eq:eta_h} and~\eqref{eq:axion_EB} through the following formula:
\begin{equation} 
    \theta \approx \frac{\pi}{\alpha \mu_0 c} \eta. \label{eq:axion_eta_rel_SA} 
\end{equation}
The effective axion field, $\theta_{\text{eff}}$, and $\eta_{\text{eff}}$ can be evaluated for the $\psi^\pm$ eigenmodes from the $\phi = 2\pi/3$ twisted cavity simulation, along with the mode's $\mathscr{H}_i$, with results shown in 
Fig.~\ref{fig:phi_axion_2}(a). In this figure, $\theta_{\text{eff}}$ and $\eta_{\text{eff}}$ are indeed proportional to each 
other for a given $\mathscr{H}_i$.

Therefore, we can state that for the twisted electromagnetic resonators, the act of twisting is 
equivalent to introducing some non-zero axion field $\theta_{\text{eff}}$ to the radiation.

\section{Discussion}

\noindent We can analytically derive the equivalency of the chirality material parameter $\kappa$ and an axion field $\theta$ at any given
point in the resonator by comparing the coefficients in the constitutive relations~\eqref{eq:CE_kappa} and~\eqref{eq:CE_axion}. 
This relationship is given by:
\begin{equation}
    \kappa = -\frac{\alpha \mu c}{\pi} \theta,\label{eq:kappa_theta}
\end{equation}
where the previously established relationship between $\eta$ and $\theta$~\eqref{eq:mixing_angle} allows us to further derive that $\kappa$ is related to $\eta$ by:
\begin{equation}
    \kappa = -\tan \eta.
    \label{eq:kappa_eta}
\end{equation}
Additionally,~\eqref{eq:eta_h} connects mixing angle $\eta$ to the helicity density $h$ at any given point in the volume of the resonator. 
Equivalent forms of these relationships for the global effective parameters $\kappa_\text{eff}$, $\theta_{\text{eff}}$, and $\mathscr{H}$ can be attained by integrating their local forms over the volume 
of the resonator. Thus they can be written as: 
\begin{equation} 
\begin{aligned}
    \eta_{\text{eff}} =& \int \eta dV,  \\
    \kappa_\text{eff} =& \int \kappa dV, 
    \label{eq:global} \\
\end{aligned}
\end{equation}
and we can see that by taking volume integrals of both sides of Equation (\ref{eq:kappa_theta}) will define $\theta_\text{eff}$.
The theoretical links between $\kappa$, $\theta$, $\eta$, and $h$ are consistent with the presented simulation results.

Our results show that twisting the boundary conditions of the cavity resonator introduces the same coupling mechanism as a non-zero, isotropic  
chirality material $\kappa$ ($\kappa_\text{eff}$) or $\theta_\text{eff}$, which mixes the $\vec{\mathbf{E}}$ and $\vec{\mathbf{H}}$ fields, leading to the 
generation of non-zero electromagnetic helicity $\mathscr{H}$ in vacuum. 
Figure~\ref{fig:phi_axion_2}(b) illustrates the relationship between these coupling mechanisms for the $\psi^\pm$ eigenmodes obtained from the $\phi = 2\pi/3$ twisted cavity simulation. 
Introducing either a twist to the boundary conditions, $\phi$, or a non-zero effective chirality to the cavity volume, 
$\kappa_\text{eff}$, results in the generation of an effective mixing angle, $\eta_\text{eff}$, between the electric 
field, $\vec{\mathbf{E}}_p$, and the magnetic field, $\vec{\mathbf{H}}_p$, of the resonant electromagnetic modes. This, in turn, leads to the emergence of an effective axion field, $\theta_\text{eff}$.
Both mechanisms converge to similar values of $\theta_\text{eff}$ and $\eta_\text{eff}$ for sufficiently large $\kappa_\text{eff}$ and $\phi$ ($\kappa_\text{eff} > 0.075$ and $\phi > \frac{5\pi}{4}$).








Each of these mechanisms introduces magneto-electric coupling that mixes the electric and magnetic fields by a mixing angle $\eta$, 
resulting in a chiral electromagnetic field. The chirality is characterised by the helicity $\mathscr{H}$, and the measurable effect of 
varying any of these three parameters is observed as a frequency shift in previously degenerate, orthogonal mode pairs. This frequency shift has been shown experimentally to agree very well with predictions.

Our approach provides a method to generate helical electromagnetic radiation in a monochromatic resonant mode in vacuo, which has potential 
applications in the measurement of material chirality. Moreover, if the twist angle of the system can 
be dynamically controlled, it is possible to actively manipulate and even reverse the electromagnetic helicity $\mathscr{H}$, opening avenues 
for applications in material diagnostics, encryption, and communication.

\section*{Acknowledgments}

\noindent This work was funded by the Australian Research Council Centre of Excellence for Engineered Quantum Systems, CE170100009 and  Centre of Excellence for Dark Matter Particle Physics, CE200100008. E.C.P is partially funded through the Defence Science Centre Research Higher Degree Student Grant.

This work used the NCRIS and Government of South Australia enabled Australian National Fabrication Facility - South Australian Node (ANFF-SA).

\section*{Availability of Data}
\noindent All data and FEM model parameters are available upon request to the authors. 

\bibliographystyle{unsrt}
\bibliography{Electromagnetic_Helicity_Twisted_arXiv_V2} 

\section*{Appendix}

\subsection{Chirality of Cr$_2$O$_3$}\label{section:chirality_parameter}

\noindent Whilst the chirality parameter $\kappa$ is purely real, there is a purely imaginary version of this parameter called the 
four-dimensional relativistic invariant pseudoscalar $\widetilde{\alpha}$. This parameter has been measured in Cr$_2$O$_3$ to be approximately 
$1.035$ $\mathrm{ps} \mathrm{m}^{-1}$ at 285 K~\cite{Hehl_Relativistic_2008}. We can derive the relationship between the two chirality parameters to be 
\begin{equation}
    \kappa=c\widetilde{\alpha}, \label{eq:kappa}
\end{equation}
by comparing the following two 
known equations for $\textbf{B}$ in the context of 
magnetoelectric media~\cite{Electromagnetic_Waves_Textbook_1994,Magnetoelectric_Effect_Fiebig_2005}
\begin{equation}
    \textbf{B}=\frac{j \kappa}{c} \textbf{E}+\mu \textbf{H},
\end{equation}
and 
\begin{equation}
    \textbf{B}=\textbf{M}+\mu_0 \mu_i \textbf{H}+\alpha \textbf{E}. 
\end{equation}
Using~\eqref{eq:kappa} we can calculate the chirality parameter $\kappa_\text{eff}$ to be $3.11\cdot 10^{-4}$ for Cr$_2$O$_3$ at 285 K. 

\subsection{Square Cross-section Resonator}

\begin{figure}[b]
    \centering
    \includegraphics[width=1\columnwidth]{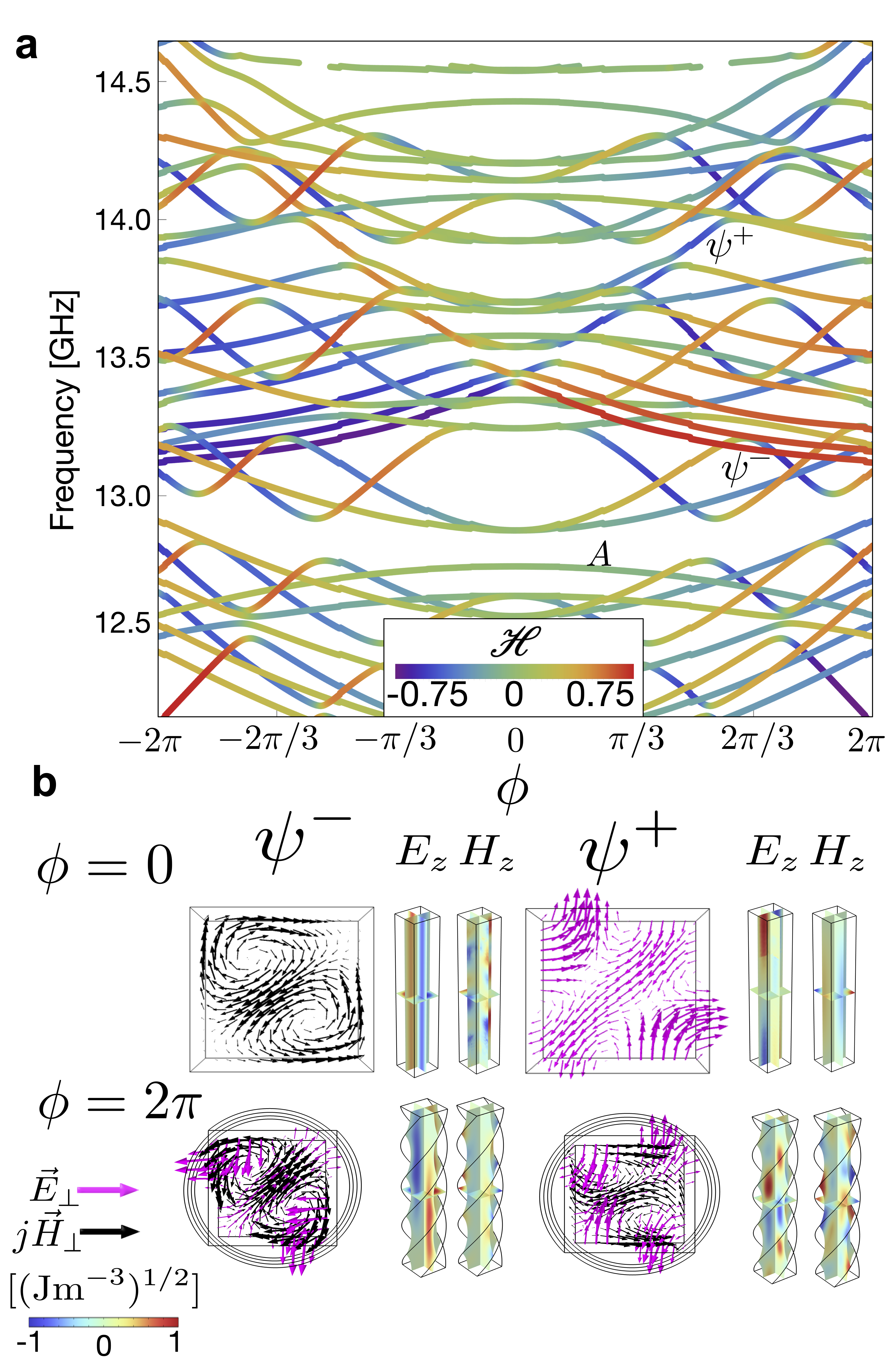}
    \caption{The eigenfrequencies of the resonant modes in the square cross-section resonator with a side length of $25$ mm as a function of $\phi$ where colour corresponds to $\mathscr{H}$.
    (b) The $\vec{E}_\perp$ (black), $j\vec{H}_\perp$ (red) fields, and corresponding energy density of the axial fields $E_z$ \& $H_z$,
    for the (a) TM$_{110}$ \& (b) TE$_{111}$ modes for the untwisted triangular cross-section resonator, and 
    the $\psi^+$ \& $\psi^-$ modes for the $\phi=2\pi$ twisted resonator.}
    \label{fig:square_mode_origin}
\end{figure}

\noindent For the resonator with the symmetric D$_4$ pentagon (i.e. square), the degenerate modes that mix together with the introduction of a 
twist $\phi$ are TE$_{20p}$ \& TM$_{21p}$ (see \ref{fig:square_mode_origin}(b)). 
The frequency tuning of the eigenmodes as a function of $\phi$ can be seen in Fig.~\ref{fig:square_mode_origin}(a) where, just like for the triangular case, $\mathscr{H}$ is generated 
with the introduction of a $\phi$, but doesn't saturate at as high a value. This is because the square case is a closer approximation to a cylindrical resonator and hence twisting has 
less of an effect on the mode mixing. The generation of the new mode orthogonality basis is shown in Fig.~\ref{fig:square_mode_origin}(b). 

\subsection{Symmetry of the Resonator Cross-Section}~\label{ap:symmetry_cross_section}
\begin{figure*}
    \includegraphics[width=2\columnwidth]{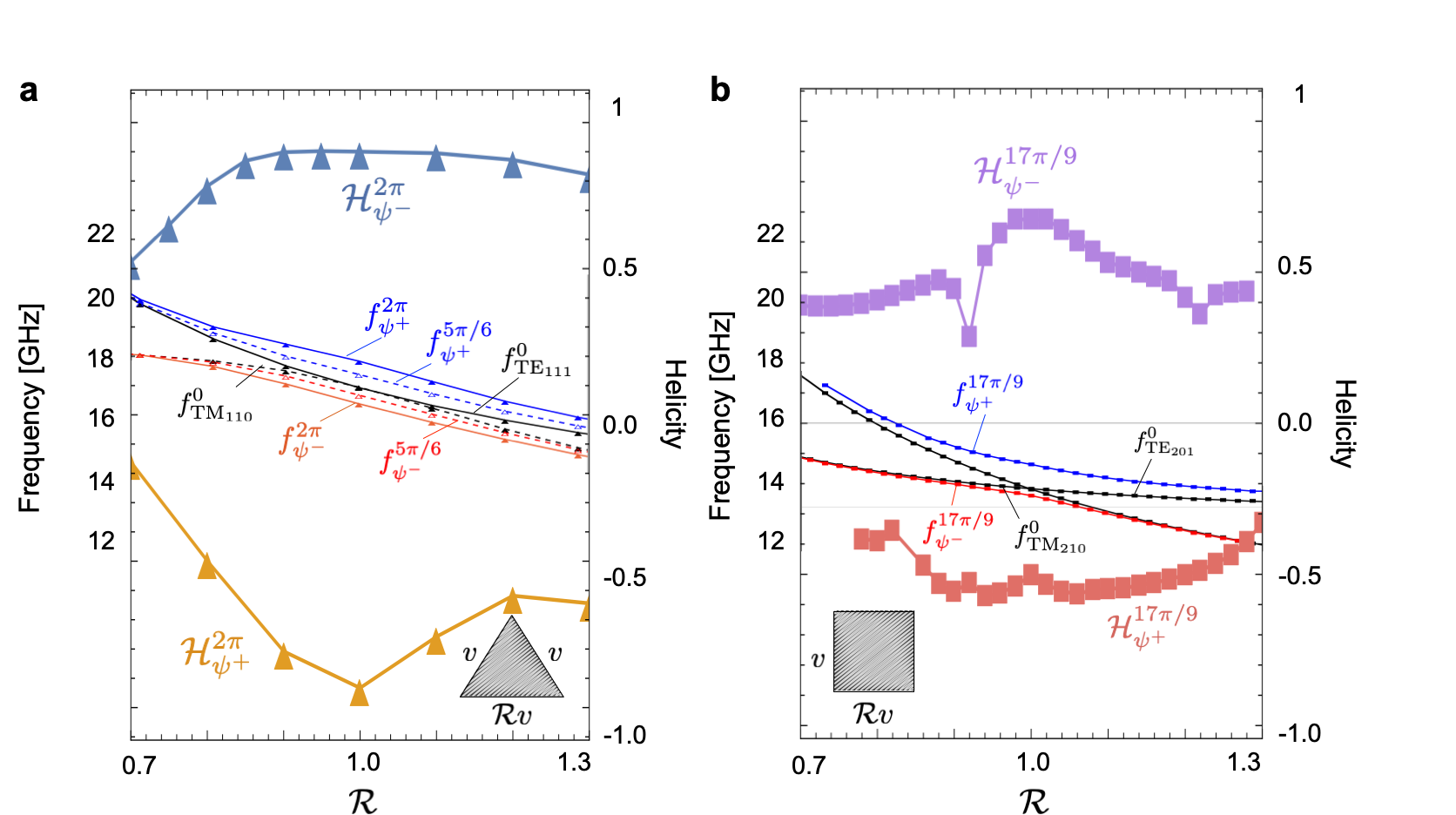}
    \caption{(a) The effect on $\mathscr{H}$ and $f_i$ as the vertex lengths of the triangular cross-section
    is multiplied by some factor $\mathcal{R}$ for different $\phi$ values. For the untwisted resonator, the $\mathrm{TE}_{111}$ and 
    $\mathrm{TM}_{110}$ $f_i$ are black \& black-dashed respectively, and for $\phi=\frac{5\pi}{6}$ and $2\pi$, 
    the $\psi^{+}$ and $\psi^{-}$ $f_i$ are blue and red respectively. For $\phi=2\pi$, $\mathscr{H}$ 
    of the $\psi^{+}$ \& $\psi^{-}$ states are gold and navy-blue respectively. 
    (b) The effect on $\mathscr{H}$ and $f_i$ as the vertex lengths of the width of the square cross-section is multipled by some factor $\mathcal{R}$ for different $\phi$ values. 
    For the untwisted square cross-section resonator, the $\mathrm{TE}_{201}$ \& $\mathrm{TM}_{210}$ modes $f_i$ are black. 
    For $\phi=\frac{17\pi}{9}$, the $\psi^\pm$ $f_i$ are blue \& red respectively, and $\mathscr{H}$ is light red \& purple respectively. }
    \label{fig:Aspect_Ratio_Sweep}
\end{figure*}

To optimise the mode mixing effect and hence $\kappa$ generation, the polygon of the resonator cross-section must be symmetric, (i.e. equilaterial triangle, square, etc.), 
which is known as the dihedral group of regular polygons. This is because this symmetry tunes the TE and TM modes closest in frequency, and for a 
coupled mode system, hybridisation of the two individual 
systems is greatest when their frequencies are made degenerate. This is demonstrated in Fig.~\ref{fig:Aspect_Ratio_Sweep}(a) \& (b) where one of the vertex
lengths of the triangular cross-section and the width of the square cross-section is 
multiplied by some factor $\mathcal{R}$. For both cases, $\mathscr{H}$ is maximised when $\mathcal{R}=1$, corresponding to the symmetric case of the respective cross-section. 

\end{document}